# Solar illumination control of ionospheric outflow above polar cap arcs

L. Maes[1], R. Maggiolo[1], J. De Keyser[1], I. Dandouras[2,3], R. C. Fear[4], D. Fontaine[5], and S. Haaland[6,7]

[1]Belgian Institute for Space Aeronomy, Brussels, Belgium, [2]Institut de Recherche en Astrophysique et Planétologie, University of Toulouse, UPS-OMP, UMR 5277, Toulouse, France, [3]CNRS, IRAP, Toulouse, France, [4]Department of Physics and Astronomy, University of Leicester, Leicester, UK, [5]Laboratoire de Physique des Plasmas, Vélizy, France, [6]Max-Planck Institute for Solar Systems Research, Katlenburg-Lindau, Germany, [7]Department of Physics and Technology, University of Bergen, Bergen, Norway

**Abstract** We measure the flux density, composition, and energy of outflowing ions above the polar cap, accelerated by quasi-static electric fields parallel to the magnetic field and associated with polar cap arcs, using Cluster. Mapping the spacecraft position to its ionospheric foot point, we analyze the dependence of these parameters on the solar zenith angle (SZA). We find a clear transition at SZA between ~94° and ~107°, with the $O^+$ flux higher above the sunlit ionosphere. This dependence on the illumination of the local ionosphere indicates that significant $O^+$ upflow occurs locally above the polar ionosphere. The same is found for $H^+$, but to a lesser extent. This effect can result in a seasonal variation of the total ion upflow from the polar ionosphere. Furthermore, we show that low-magnitude field-aligned potential drops are preferentially observed above the sunlit ionosphere, suggesting a feedback effect of ionospheric conductivity.

## 1. Introduction

The polar ionosphere is a major source of outflowing ions [*Hultqvist et al.*, 1999; *Yau and André*, 1997]. Because these ions flow on open field lines, they can escape into interplanetary space and contribute to atmospheric erosion [*Haaland et al.*, 2012; *André and Cully*, 2012]. If trapped in the magnetosphere, they can affect magnetospheric dynamics due to the significant amount of heavy ions (mostly $O^+$) in the ionospheric plasma [*Lotko*, 2007; *Kronberg et al.*, 2014]. Therefore, it is important to have a correct assessment of the flux and composition of outflowing ions above the polar caps.

One main outflow mechanism is the polar wind [see *Yau et al.*, 2007, and references therein]. Due to the open geometry of the magnetic field, no hydrostatic equilibrium can be established so that ionospheric plasma can escape [*Dessler and Michel*, 1966]. The lighter electrons tend to escape more easily than the ions, which causes a charge separation that sets up an ambipolar electric field that decelerates the electrons and accelerates the ions to guarantee neutral outflow [*Axford*, 1968; *Banks and Holzer*, 1968]. This electric field maintains a constant flow of plasma with energies of a few eV from the ionosphere into the magnetospheric lobes [*Engwall et al.*, 2009]. In the classical polar wind theory, $O^+$ ions are deemed too heavy to escape. Yet observations have shown a significant amount of $O^+$ in the polar wind [*Nagai et al.*, 1984; *Waite et al.*, 1985; *Abe et al.*, 1993; *Su et al.*, 1998a]. Therefore, additional acceleration mechanisms have been proposed (see *Tam et al.* [2007] for an overview).

A second high-latitude escape mechanism is cusp outflow. Soft electron precipitation and wave activity [e.g., *Zheng et al.*, 2005; *Moore and Khazanov*, 2010; *Nilsson et al.*, 2012] heat the ionosphere below the cusp and energize the escaping ions [*Lockwood et al.*, 1985]. This way, $O^+$ ions can be energized enough to overcome the gravitational potential. Despite the cusp's small spatial extent, cusp outflow is a significant source of magnetospheric ions, with total number fluxes estimated at $10^{25}$ $s^{-1}$ [*Nilsson et al.*, 2012] and with energies typically higher than those of polar wind ions. It has been suggested that the $O^+$ ions observed in the polar wind actually originate in the cusp and have drifted over the polar cap [e.g., *Green and Waite*, 1985; *Nilsson et al.*, 2012], eliminating the necessity for the additional acceleration mechanisms mentioned earlier.

A third ion escape route is through acceleration by quasi-static electric fields parallel to the magnetic field and associated with polar cap arcs. While similar to discrete arcs in the auroral oval, polar cap arcs occur poleward of the auroral oval. They typically appear during quiet times [*Davis*, 1963] and periods of prolonged northward interplanetary magnetic field (IMF) [*Berkey et al.*, 1976]. In this paper we focus on small-scale polar cap arcs, as opposed to the larger-scale theta auroras (see *Zhu et al.* [1997] for an overview of classification schemes).

At high altitudes above polar cap arcs, in situ measurements detect upward accelerated ion beams characterized by strongly field-aligned velocities [*Maggiolo et al.*, 2006, 2012]. There is a significant amount of $O^+$ ions present in the upflow, although the $H^+$ ions mostly dominate [*Maggiolo et al.*, 2011]. In contrast to cusp outflow, where the ratio of $O^+$ over $H^+$ energies is well above unity (typically ∼4 to ∼16 when detected by Cluster above the polar ionosphere [*Maggiolo et al.*, 2006; *Nilsson et al.*, 2012]), the energy of both species in polar cap ion beams is roughly the same as a consequence of the electrostatic acceleration of the ions by a strong field-aligned electric field. This parallel electric field is part of a U-shaped potential profile associated with a bipolar perpendicular electric field structure at high altitude [*De Keyser and Echim*, 2010]. Integrating the high-altitude perpendicular electric field correlates well with the maximum ion energy [*Maggiolo et al.*, 2006, 2011]. This parallel electric field at the same time accelerates electrons downward, which causes the luminescence of the polar cap arc [*Maggiolo et al.*, 2012].

Similar potential structures have been observed in discrete auroral arcs and have been studied extensively [e.g., *Lyons et al.*, 1979; *Akasofu*, 1981; *Ergun et al.*, 1998; *Marklund et al.*, 2011]. The field-aligned potential drops above polar cap arcs, 400 V on average [*Maggiolo et al.*, 2011], are smaller than those associated with discrete arcs in the auroral oval, which typically are a few kV [*Partamies et al.*, 2008].

The objective of this study is threefold. The first goal is to measure the composition of the outflowing ion beams above polar cap arcs. Second, we register the energy of the upward accelerated ions as a proxy for the magnitude of the field-aligned potential drop. Finally, we try to assess the importance of solar illumination of the underlying ionosphere on the ion upflow. The paper is organized as follows. After a description of the data, the method, and the event selection, the results of a statistical analysis are presented. We conclude with a discussion of these results in terms of the role of solar illumination, and we infer their implications for the total ionospheric upflow and outflow.

## 2. Data and Method

Early in the mission, the European Space Agency's (ESA) Cluster satellites were particularly well suited to study polar cap ion beams, as the constellation of four spacecraft was in a 4 $R_E$ × 19.6 $R_E$ quasi-polar orbit [*Escoubet et al.*, 1997]. Plasma composition measurement is the goal of the Composition and Distribution Function analyser (CODIF), part of the Cluster Ion Spectrometry experiment (CIS), which can differentiate between ions with different masses and energies up to 38 keV [*Rème et al.*, 2001]. The other instrument used here is the Hot Ion Analyser (HIA), also part of CIS. With an energy range from ∼5 eV to 32 keV, HIA has a larger sensitivity than CODIF but cannot distinguish between different ion species [*Rème et al.*, 2001]. We therefore extract the $O^+$ and $H^+$ densities in the ion beams from the CODIF data, while we obtain the energy of the ions from HIA.

The sample of ion beams used here is a selection from the database of ∼200 events discussed by *Maggiolo et al.* [2011]. In about 40% of these events a hot isotropic background population is present in addition to the upflowing ions. This 40% is not included in our set because in such cases it is not always possible to clearly separate both populations. Also, in order to avoid possible intercalibration issues, we only use measurements from CODIF on Cluster 1, until September 2004, when that instrument stopped working. Events were also eliminated if there were not enough data to obtain a decent median density (i.e., less than four data points). The resulting sample contains 67 events, relatively well distributed over local time and latitude.

The events were characterized by considering the central 60% of the time interval during which the beams are seen. Indeed, due to the U-shaped potential the ion energy is lower at the beam edges, where the measurements can suffer from the effects of the spacecraft potential that causes part of the ion population to be missed by the detector. In addition, the perpendicular electric field at the edges may also cause ions to drift which can result in significant horizontal transport away from their origin in the local ionosphere. The energy of maximum of flux is taken from the HIA data in the center of the inverted V where the upward acceleration is maximum.

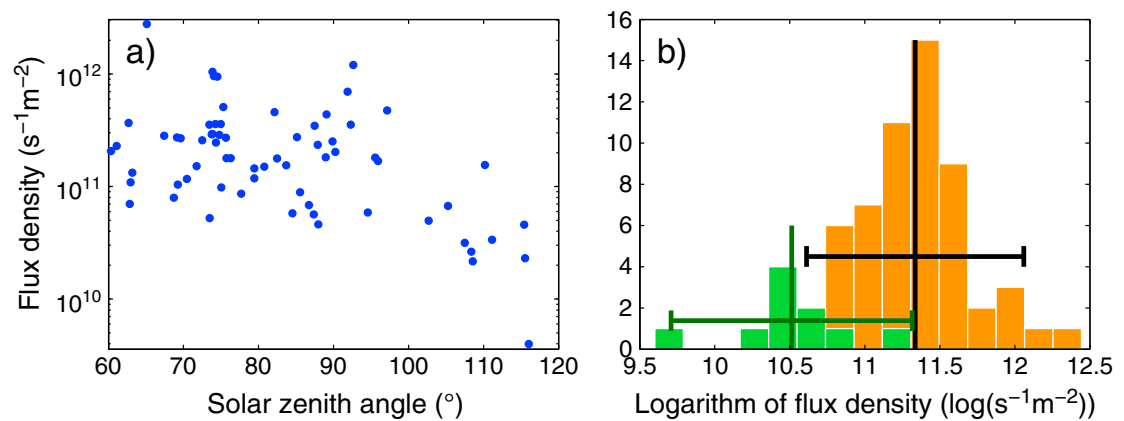

**Figure 1.** (a) The O$^+$ flux densities versus the SZA of the ionospheric foot point. (b) The histograms of the logarithm of the O$^+$ flux densities. In orange (green) for the events with SZA lower (higher) than 100°. The vertical black (green) line gives the mean of the orange (green) distribution; the horizontal lines show the 1.96σ interval.

Because these are cold ion beams, the measured energy can be used to calculate the field-aligned ion velocity. Multiplying this with the O$^+$ and H$^+$ densities, we obtain an estimate for the flux densities. If the beams are exactly field aligned, the cross-sectional area scales inversely to the magnetic field strength; this property is exploited to normalize the flux densities to a common ionospheric altitude of 200 km. By virtue of flux conservation, this provides a measure of the upward ion flux at the altitude of the bottom of the acceleration region. This altitude is not precisely known yet; for discrete auroral arcs observations find it to be between ∼0.5 and ∼2 $R_E$ [see *Karlsson*, 2012, and references therein], with simulations showing that the altitude of the acceleration region depends on the energy: at higher altitudes for lower energies [e.g., *Gunell et al.*, 2013]. For polar cap arcs, with potential drops on average an order of magnitude smaller than for auroral arcs, this will thus likely be around ∼1 $R_E$ altitude or higher.

To trace the origin of the upflowing ions, we use the T96 magnetic field model [*Tsyganenko and Stern*, 1996] to map the spacecraft position back along a magnetic field line down to its foot point in the ionosphere at 200 km altitude. We then calculate the solar zenith angle (SZA) at that foot point, which is the angle between the normal to the surface and the direction to the Sun, since this reflects the solar illumination conditions in the local ionosphere. The tracing of magnetic field lines depends on a magnetic field model and could thus introduce some error. At the altitude of the Cluster spacecraft, however, the total magnetic field is still dominated by Earth's internal magnetic field so that our estimations of these errors are smaller than 1°.

## 3. Results

Figure 1a displays the measured O$^+$ number fluxes, ordered according to the SZA of the corresponding field line foot point in the ionosphere. The O$^+$ flux densities range over almost 3 orders of magnitude, from $4.0 \times 10^9$ to $2.8 \times 10^{12}$ m$^{-2}$ s$^{-1}$, with a mean of $2.8 \times 10^{11}$ m$^{-2}$ s$^{-1}$. The figure shows a transition from higher flux densities to lower ones between ∼94° and ∼107° SZA; the precise location of the transition is hard to pinpoint. Events below 94° SZA all have O$^+$ flux densities higher than $4.6 \times 10^{10}$ m$^{-2}$ s$^{-1}$, with a mean of $3.3 \times 10^{11}$ m$^{-2}$ s$^{-1}$, while those above 107° SZA are all lower than $1.5 \times 10^{11}$ m$^{-2}$ s$^{-1}$, with a mean of $4.3 \times 10^{10}$ m$^{-2}$ s$^{-1}$.

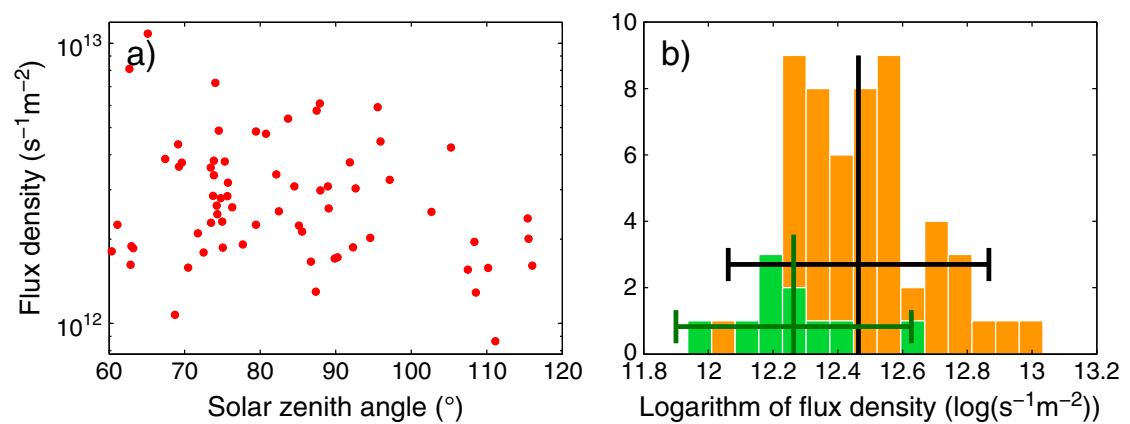

**Figure 2.** (a) The H$^+$ flux densities versus the SZA of the ionospheric foot point. (b) The histograms of the logarithm of the H$^+$ flux densities. In orange (green) for the events with SZA lower (higher) than 100°. The vertical black (green) line gives the mean of the orange (green) distribution; the horizontal lines show the 1.96σ interval.

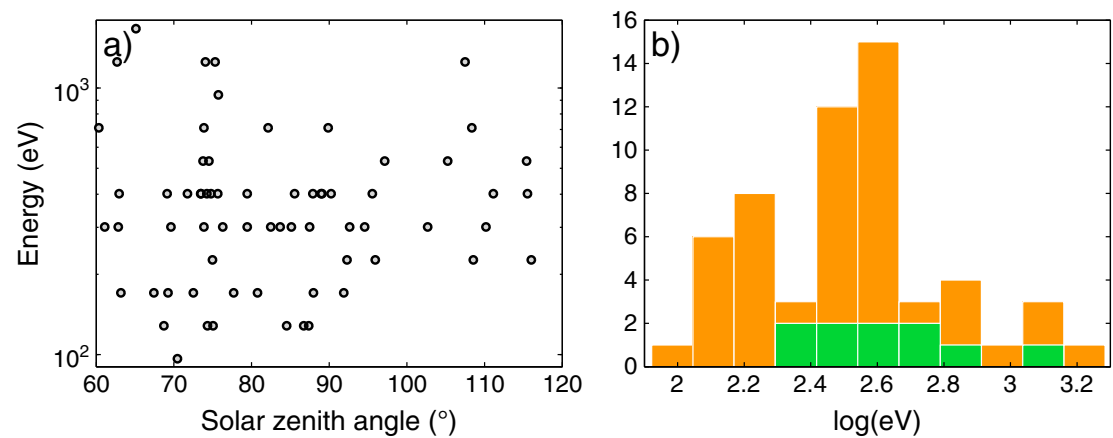

**Figure 3.** (a) The energy of maximum of flux in the ion beams versus the SZA of the ionospheric foot point. (b) The histograms of the logarithm of the energy. In orange (green) for the events with SZA lower (higher) than 100°.

To check the statistical significance of this transition, the events are divided into two groups: below and above 100° SZA. The distributions of the logarithms of the flux densities are given for both groups in Figure 1b. Note that the mean of the set above 100° SZA falls outside the 1.96$\sigma$ interval of the set below 100° SZA; in fact, it is even smaller than the lowest value in the set below 100° SZA. In the same way, the mean of the < 100° distribution is higher than the highest flux density of the > 100° distribution. The statistical significance of the difference between both sets is quite high. When performing a *t* test, we find a probability of $7.5 \times 10^{-5}$ that both sets were drawn from the same lognormal distribution of number fluxes (a reasonable choice given the nature of the distributions in Figure 1b). Using a Mann-Whitney-Wilcoxon test, which does not require normality, we find this probability to be $5.4 \times 10^{-6}$.

In Figure 2a we plot the H$^+$ flux densities versus the SZA of the foot points. With values between $8.6 \times 10^{11}$ and $1.1 \times 10^{13}$ m$^{-2}$ s$^{-1}$, and with a mean of $3.1 \times 10^{12}$ m$^{-2}$ s$^{-1}$, the flux density is on average an order of magnitude higher than the O$^+$ flux density and ranges over a bit more than 1 order of magnitude. Here too we find a transition between ~94° and ~107°, with slightly lower values at higher SZA. However, the difference is smaller, with mean values of $3.2 \times 10^{12}$ m$^{-2}$ s$^{-1}$ below and $1.7 \times 10^{12}$ m$^{-2}$ s$^{-1}$ above the transition so that the change is less clear than for O$^+$. This is also seen in the histogram of the logarithm of the H$^+$ number flux in Figure 2b. The distributions found for SZA below and above 100° differ less. The mean of the set with SZA > 100° is well within the 1.96$\sigma$ interval of the set with SZA < 100°. The probability that both sets come from the same distribution is higher, $8.3 \times 10^{-3}$ from the *t* test and $5.3 \times 10^{-3}$ from the Mann-Whitney-Wilcoxon test, but the division in two distinct sets still has a high statistical significance.

The energy of maximum ion flux is shown in Figure 3a, ordered according to SZA. The minimum beam energy is 96 eV, the maximum is 1669 eV, and the mean energy is 422 eV. Interestingly, we again see a change in the distribution of energies around ~100°. Here also, it is not clear where exactly the transition occurs. The difference between the mean energies, 410 eV below and 489 eV above 100°, is not large, but at the higher angles no event with an energy lower than 227 eV is observed, whereas for the lower angles 26% of the observed events have energies below this value. In other words, the first quartile (i.e., the lowest 25%) of the events at SZA below 100° is 170 eV, which is lower than the lowest energy of the events above 100°. On the other hand, the third quartile at the lower SZA is 410 eV, and there are four events (out of 10) at the higher SZA with energies higher than this.

## 4. Discussion

We have found a statistically significant difference in the fluxes of outflowing O$^+$ and H$^+$ above polar cap arcs, as well as in their energies, below and above a transition in solar zenith angle that separates a sunlit and a dark ionosphere. Therefore, the data clearly suggest a modulation of the upflow from the ionosphere above polar cap arcs by solar illumination.

The transition is found between ~94° and ~107°. The SZA of the terminator at an ionospheric altitude *h* is given by $h = R \cdot (1/\cos(\text{SZA} - 90°) - 1)$, where *R* is the Earth's radius augmented by the altitude where the atmosphere becomes opaque to UV light at the relevant wavelengths, i.e., the altitude of the ozone layer of ∼30 km (assuming that the atmosphere is optically thin above and completely opaque below that altitude). This nonlinear relation between $h$ and SZA implies that the terminator at an altitude of ∼45 km has a SZA = 94°, at ∼130 km it is 100°, and at ∼310 km it is 107°. The observed modulation therefore suggests that the $O^+$ source is situated at the ionospheric field line foot point directly below the observed beams. It also places this source somewhere between this ∼130 km and ∼310 km, which is fully compatible with the fact that $O^+$ ionization in the ionosphere is most efficient in the $F$ layer above 150 km altitude. It is also known that the $F$ layer is depleted during nighttime, i.e., that there is a strong solar illumination effect on the ionization in the $F$ layer.

The solar illumination effect leads to a modification of the composition in the ion beams. The $O^+/H^+$ number flux ratio in the ion beams above a sunlit polar cap, 0.095, is much larger than above a dark polar cap, where it is only 0.026. In terms of mass flux, due to their larger mass, oxygen ions are dominant on the sunlit side, with the ratio being 1.5, but dropping to 0.41 on the dark side. An obvious explanation for this difference in ion outflow could be the observed day/night difference in the $O^+$ ion content in the $F$ layer: There simply are more ions available to flow up during daytime. Composition and density changes in the $F$ layer will result in changes at higher altitudes. The presence of photoelectrons could also affect the ion upflow above the polar cap, as suggested by several papers [e.g., *Axford*, 1968; *Tam et al.*, 1995, 1998; *Glocer et al.*, 2012]. An increased scale height in the $F$ region ionosphere will facilitate the $O^+$ ions reaching the bottom of the acceleration region and thus is also an important factor in the control of the ion upflow.

The time it takes for the ionosphere to react to the absence of solar illumination after having been sunlit (i.e., relaxation time), or to the presence of solar illumination after having been in the dark, could affect the results, blurring the difference between the sunlit and the dark side of Figures 1a and 2a. However, when we assume the terminator to be at a SZA of, e.g., 100°, we find that all events except one have foot points in the ionosphere on positions that have been in or out the sunlight for more than 4 h; most have even been there for more than a day.

Some authors have suggested that upflowing $O^+$ ions above the polar cap originate in the cusp and have drifted across the polar cap [e.g., *Green and Waite*, 1985; *Nilsson et al.*, 2012]. However, the observed dependence of the properties of the polar cap ion beams on the SZA at the field line foot point rather suggest that the $O^+$ ions we observe in the ion beams originate in the local polar ionosphere. We also do not find a correlation between the flux densities and the distance from the cusp. Further corroborating the idea of an origin in the local ionosphere is the fact that our data are collected during, or not long after, periods of northward IMF during which there is little or no antisunward convection across the polar cap. It is also unlikely that ions have entered the beams from the sides and crossed the magnetic field lines by means of some form of diffusive transport, because then we would expect a large dispersion in the energies of these ions, while we observe a rather narrow energy distribution.

It should be noted that even during northward IMF, the convection could be a cause of errors of a few degrees in determining the SZA. This would blur the transition, certainly considering that convection is not only antisunward during northward IMF. Therefore, the fact that we do see the transition means that the effect of convection is less significant than that of solar illumination.

The energy of the accelerated ions is a good proxy for the magnitude of the field-aligned potential drop. Therefore, our findings show that the potential drop associated with polar cap arcs also changes between a sunlit and a dark ionosphere. Several studies have found evidence of a dependence of the field-aligned potential drop associated with auroral arcs on solar illumination [*Newell et al.*, 1996, 2010; *Cattell et al.*, 2006; *Liou et al.*, 2011]. However, they all show the higher energies to be suppressed in the sunlight. *Liou et al.* [2011] also report a monotonic increase of the potential drop with SZA up to 108°. Note, however, that one should be cautious with the analogy with discrete arcs in the auroral oval; these may behave differently and typically involve larger potential differences [*Partamies et al.*, 2008] than those observed above the polar cap arcs in this paper and occur in regions of intense particle precipitation.

Solar illumination enhances the ionospheric electron and ion density and temperature, thereby increasing the horizontal conductivity. Since the field-aligned electric fields in quasi-static arcs are part of a current system with a generator somewhere in the magnetosphere and closing horizontally in the ionosphere, it is likely that the ionospheric conductivity can affect the whole current system and thus also the field-aligned potential drop [e.g., *Lyons*, 1980, 1981; *De Keyser and Echim*, 2010].

In the case of discrete auroral arcs, the precipitating particles dump a lot of energy into the underlying ionosphere. In polar cap arcs the precipitating energy fluxes are smaller due to the lower parallel potential drops and to the absence of trapped populations on polar magnetic field lines. The relative contribution of solar illumination in heating and ionizing the underlying ionosphere is therefore correspondingly larger. Indeed, we did not find a correlation between the flux density and the magnitude of the potential drop, indicating that the precipitating electrons do not significantly affect the ionosphere below, i.e., that the properties of the ionosphere below a polar cap arc are not significantly different from that in the rest of the polar cap.

One might therefore conjecture that the ion outflow above the polar cap ionosphere, i.e., the polar wind, is very similar to the outflow above a polar cap arc, the only difference being that the ionospheric ions above a polar cap arc are accelerated enough so that they can be reliably detected by a charged spacecraft, while the observation of the cold polar wind itself is inherently difficult due to spacecraft potential issues. Although further study is warranted, we can compare our results to studies of $O^+$ in the polar wind.

The models of *Su et al.* [1998a, 1998b] and *Glocer et al.* [2012] show a strong decrease in $O^+$ density at solar zenith angles between 95° and 105° above the polar cap when incorporating photoelectrons in the polar wind. From satellite observations, *Abe et al.* [1993] report a transition in the upward ion velocities in Akebono data, from higher on the dayside to lower on the nightside, at altitudes from 5000 km to 9000 km. Using measurements from the Polar spacecraft, *Su et al.* [1998a] find a strong decrease in $O^+$ densities and downward fluxes going from 90° to 105° at altitudes of 5000 km. They conclude that most $O^+$ ions originate in the cleft ion fountain or cusp upflow. As mentioned before, in our data we find evidence that this is not the case for the ions observed in the ion beams, which originate in the local ionosphere and have entered the acceleration region from the bottom. It is possible that this study and *Su et al.* [1998a] consider different populations. *Su et al.* [1998a] measure much more $O^+$ ions (they find $O^+$ being the dominant species at 5000 km altitude) and see mostly downflowing ions originating in the cusp. It is likely that these downflowing ions would not enter the acceleration region and that we only observe the small population of $O^+$ ions flowing up from the local ionosphere.

With the aforementioned caveats in mind, we can extrapolate the average measured fluxes to the whole polar cap area. Approximating the poleward boundary of the auroral oval with a circle at 75° magnetic latitude, and using the mean flux densities over all events, we estimate the total upward $H^+$ and $O^+$ number flux (up to the altitude of the bottom of the acceleration region) to be $2.9 \times 10^{25}$ s$^{-1}$ and $2.6 \times 10^{24}$ s$^{-1}$, respectively. Note that these values are obtained mostly during northward IMF conditions. Because of the sunlit/dark difference, we would expect daily and seasonal variations of the total ion flux above the polar cap, most notably in the $O^+$ flux. We could use the mean flux density of events at SZA > 100° for upflow from a dark polar cap and the mean of the other events for a sunlit polar cap. Then, for example, from a fully sunlit polar cap, like during summer solstice, the extrapolated total $O^+$ upflow would be around $3.0 \times 10^{24}$ s$^{-1}$, whereas from a polar cap in darkness, like around midnight during winter solstice, it would only be $4.2 \times 10^{23}$ s$^{-1}$. Because the sunlit/dark transition occurs at SZA larger than 90°, the variations in the two hemispheres do not cancel each other. The ion upflow, and thus probably also the outflow, from both polar caps combined should still exhibit a seasonal variation, namely, from solstice to equinox.

## 5. Summary

In summary, we find a strong difference in upward $O^+$ flux above polar cap arcs depending on whether the magnetic field line foot point is sunlit or not. This is also found for $H^+$ but less pronounced. The magnitude of the field-aligned potential drop is also found to behave differently. Solar illumination of the local polar ionosphere (particularly the *F* layer) is therefore considered the main parameter controlling the upflow. Furthermore, the existence of two distinct regimes for the sunlit and dark polar ionosphere suggests that a diurnal and seasonal variation for the total polar wind outflow may exist.

**Acknowledgments**

This work was supported by the Interuniversity Attraction Pole Planet TOPERS initiated by the Belgian Science Policy Office and by the ESA-Prodex Project CON-Cluster. We acknowledge the support from the International Space Science Institute through funding of their International Team on Polar Cap Arcs. The authors are grateful to E. Penou for the development of CL, the Cluster CIS software. All Cluster data can be found at the Cluster Science Archive (www.cosmos.esa.int/web/csa/). The dates and times of the events used in this study can be found in the supporting information.

The Editor thanks Charles Chappell and an anonymous reviewer for their assistance in evaluating this paper.